  \documentstyle[]{article}
\def\bea{\begin{eqnarray}}
\def\eea{\end{eqnarray}}
\def\nn{\nonumber}

\def\beq{\begin{equation}}
\def\eeq{\end{equation}}
\def\ba{\beq\new\begin{array}{c}}
\def\ea{\end{array}\eeq}
\def\be{\ba}
\def\ee{\ea}
\def\stackreb#1#2{\mathrel{\mathop{#2}\limits_{#1}}}
\def\Tr{{\rm Tr}}
\def\res{{\rm res}}

\makeatletter
\newdimen\normalarrayskip              
\newdimen\minarrayskip                 
\normalarrayskip\baselineskip
\minarrayskip\jot
\newif\ifold             \oldtrue            \def\new{\oldfalse}
\def\arraymode{\ifold\relax\else\displaystyle\fi} 
\def\eqnumphantom{\phantom{(\theequation)}}     
\def\@arrayskip{\ifold\baselineskip\z@\lineskip\z@
     \else
     \baselineskip\minarrayskip\lineskip2\minarrayskip\fi}
\def\@arrayclassz{\ifcase \@lastchclass \@acolampacol \or
\@ampacol \or \or \or \@addamp \or
   \@acolampacol \or \@firstampfalse \@acol \fi
\edef\@preamble{\@preamble
  \ifcase \@chnum
     \hfil$\relax\arraymode\@sharp$\hfil
     \or $\relax\arraymode\@sharp$\hfil
     \or \hfil$\relax\arraymode\@sharp$\fi}}
\def\@array[#1]#2{\setbox\@arstrutbox=\hbox{\vrule
     height\arraystretch \ht\strutbox
     depth\arraystretch \dp\strutbox
     width\z@}\@mkpream{#2}\edef\@preamble{\halign
\noexpand\@halignto
\bgroup \tabskip\z@ \@arstrut \@preamble \tabskip\z@ \cr}%
\let\@startpbox\@@startpbox \let\@endpbox\@@endpbox
  \if #1t\vtop \else \if#1b\vbox \else \vcenter \fi\fi
  \bgroup \let\par\relax
  \let\@sharp##\let\protect\relax
  \@arrayskip\@preamble}
\def\eqnarray{\stepcounter{equation}%
              \let\@currentlabel=\theequation
              \global\@eqnswtrue
              \global\@eqcnt\z@
              \tabskip\@centering
              \let\\=\@eqncr
              $$%
 \halign to \displaywidth\bgroup
    \eqnumphantom\@eqnsel\hskip\@centering
    $\displaystyle \tabskip\z@ {##}$%
    \global\@eqcnt\@ne \hskip 2\arraycolsep
         $\displaystyle\arraymode{##}$\hfil
    \global\@eqcnt\tw@ \hskip 2\arraycolsep
         $\displaystyle\tabskip\z@{##}$\hfil
         \tabskip\@centering
    &{##}\tabskip\z@\cr}
\newfont{\hr}{msbm10}
\newfont{\ams}{msam10}

\begin{document}


\begin{titlepage}
\setcounter{footnote}0
\begin{center}
\hfill FIAN/TD-19/95\\
\hfill ITEP/TH-4/96\\
\vspace{0.3in}
{\LARGE\bf  Exact solutions to quantum field theories and integrable equations}
\footnote{{\sl talk given at:}  {\it Singular limits of
nonlinear dispersive waves}, September 4-9 1995, Zvenigorod, Russia and
{\it Supersymmetries and quantum symmetries}, September 25-30, Dubna, Russia}
\\
\bigskip
\bigskip
\bigskip
{\Large A.Marshakov}
\footnote{E-mail address:
mars@lpi.ac.ru}\\
{\it Theory Department,  P. N. Lebedev Physics
Institute , Leninsky prospect, 53, Moscow,~117924, Russia\\
and\\ ITEP, Moscow 117259, Russia}\\
\bigskip
\bigskip
\end{center}
\bigskip \bigskip

\begin{abstract}
The exact solutions to quantum string and gauge field theories are discussed
and their formulation in the framework of integrable systems is presented.
In particular I consider in detail several examples
of appearence of solutions to the first-order integrable equations of
hydrodynamical type and stress that all known examples can be treated as
partial solutions to the same problem in the theory of integrable systems.
\end{abstract}

\end{titlepage}

\newpage
\setcounter{footnote}0

\section{Introduction}

In this note I will try to describe some recent advances in modern quantum
field theory related with the appearence of several examples of the {\it
exact quantum} solutions. Amazingly enough it turns out that an
adequate language for the description of the exactly solved quantum theories
can be found in the framework of classical integrable systems and some
essential role is played by rather simple first order equations of
hydrodynamical type or so called Whitham equations on moduli spaces
of complex structures of Riemann surfaces appearing when solving
(quasi)periodic problems for the systems of KP or Toda type.

The original problem in quantum field theory defined by the ''bare" action
functional
\beq\label{act}
S = \int d^Dx {\cal L}(\varphi, \partial _{\mu}\varphi,\dots;g_k)
\eeq
is to compute the exact quantum correlation function
\beq\label{corr}
\langle \Phi _{k_1}\dots\Phi _{k_n}\rangle = \int D\varphi e^{-{1\over\hbar}S}
\Phi _{k_1}(\varphi ,\partial _{\mu}\varphi ,\dots )\dots\Phi_{k_n}(\varphi ,
\partial _{\mu}\varphi ,\dots ) \equiv F_{k_1,\dots ,k_n}(g,u)
\eeq
as a function of coupling constants $\{g_k\}$, quantum characteristics of
physical operators $\{k_1,\dots ,k_n\}$ (in general, of course, not being
just a discrete set of numbers) and possible parameters of classical solutions
$\{ u_{\alpha}\} $ or {\it moduli} of the theory.

Unfortunately almost never the integral in (\ref{corr}) can be computed
exactly. Up to now the only working methods in quantum field theory are given
by perturbation theory and lattice regularization of the problem (\ref{corr})
resulting usually in some approximations being quite far from the exact answer. However,
sometimes it turns out that it is possible to propose the exact answer
not computing the path integral directly and there exists a beleif that
such answers are nevertheless true.

First, this is the case of some models of string theory. Using an
elegant discretization which is quite far from an original $D=2$ Lagrangian
it has been possible to compute exact solutions to the nonperturbative
quantum string correlators in the technique of matrix models. An important
point is that the answer was proposed in the form of a solution to
the hierarchy of classical integrable equations.
The most
advanced results were obtained for so called topological $2D$ field
theories where it has turned to be possible to get the solutions explicitely.
\footnote{There exists a big set of review papers devoted to this subject.
In \cite{mamo} only those where the approach to $2D$ problems is mostly
closed to that developed below are presented
(see also references therein).}

Second, there are examples of the exact solutions in $4D$ supersymmetric
gauge theories \cite{WS,VaWi} deduced from the investigation of duality
properties and an assumption of complex analiticity of quantum answers
as functions of coupling constants and/or moduli parameters. An important
observation is \cite{WSGP} that the proposed exact answer is again a
particular solution to certain differential equation.

Below I will try to pay attention to this {\it universal} property of
the exact quantum field theory solutions. It will be demonstrated that
various exact quantum answers to {\it different} theories can be
obtained as different solutions almost to  {\it the same} classical
problem of (integrable) differential equations. Unfortunately it is still
unclear what is the concrete mechanism of arising these equation in quantum
field theory.

The main hope is that the (first - order) equations of hydrodynamical type
are generalizations of the renormalization-group technique of the
perturbation theory. Indeed it is known that in the perturbation theory the
scale dependence of quantum correlation functions is related with their
dependence on the coupling constants by means of the first order equation
\beq\label{rg}
\left( {d\over d\log\mu} - \sum \beta _i(g){\partial\over\partial g_i}\right)
F(g;\mu ) = 0
\eeq
Naively in the exact solution there is no scale dependence, so the
resulting equation could be of the form $\sum \beta _i(g){\partial F\over
\partial g_i} = 0$, or the derivatives over $\mu $ could be replaced by the
derivatives over moduli $\{ u_{\alpha }\}$. Even accepting this it is necessary
to stress that there is no known way to define $\beta $-function beyond the
perturbation theory (i.e. not using the scale dependence $\beta _i(g) =
{\partial g_i\over \partial\log\mu}$). One can also hope to find integrable
structures in the nonperturbative calculations investigating the moduli
spaces of classical solutions (like instantons in gauge theories or $2D$
$\sigma $-models) and this way may also lead towards the Hitchin-type
constructions \cite{Hi}.

In what follows I will concentrate mostly on two examples of quantum exact
solutions. The first one is the simplest case of $D=2$ topological string
theory -- pure topological gravity where the exact generating function of
all quantum correlators is known \cite{Kontsevich}. In this example the
correlators
$F_{k_1\dots k_n}$ are just integer numbers -- so called intersection
indices on moduli spaces of complex structures of Riemann surfaces.
The second (almost a realistic)  example
is supersymmetric Yang-Mills theory. It is not yet possible to get all quantum
answers in the
latter case but what was proposed recently in \cite{WS} is the exact
nonperturbative
dependence of the coupling constants $\tau = \theta + {i\over g^2} = \tau (u)$
where (the bare) constants can be read off the Lagrangian
\beq\label{ym}
{\cal L} = {1\over g^2}\Tr F_{\mu\nu}^2 + i\theta \Tr F_{\mu\nu}
{\tilde F}_{\mu\nu} + \dots
\eeq
on the moduli of the exact solution $\{ u_{\alpha }\} $ -- vacuum expectation
values of
the Higgs field as well as the allowed massive spectrum of the theory
\beq\label{mass}
M = \left|{\bf n a}(u) + {\bf m a}_D(u) \right| =
\left| \left( {\bf n}\oint _{\bf A} + {\bf m}\oint _{\bf B}\right)dS \right|
\eeq
and the effective action of the ''light modes" determined by
\be\label{f}
a_{D,i} = {\partial {\cal F}\over\partial a_i} \ \ \  \ \ \ \
\tau _{ij}= {\partial a_{D,i}\over\partial a_j} = {\partial ^2 {\cal F}
\over\partial a_i\partial a_j}
\ee
Now we will see that the above formulas can be considered
as different solutions of the same problem concerning classical integrable
equations.

There are two main questions concerning the appearence of the integrable
equations. First one -- how the effective action of a quantum field theory
becomes related with the fundamental object in the theory of integrable systems
${\cal F} = \log {\cal T}$ -- the
$\tau $-function of an integrable hierarchy and what is the mechanism
expressing the data (\ref{mass}) in terms of
the generating differential $dS$ of the {\it Whitham} hierarchy \cite{WSGP}.
The second question arises since the Whitham equations have in principle lots
of solutions -- so what is the way to distinguish the ''physical" ones
appearing
in the vicinity of a finite-gap solutions to integrable systems of a
KP/Toda-type. Therefore, one should
learn what kind of spectral curves (i.e. the finite-dimensional integrable
systems) arise in the context of quantum field
theory and discuss the particular solutions determined by these spectral
curves.

\section{Whitham equations}

Let us briefly remind the most general definition of the
Whitham hierarchy given by Krichever \cite{Kri}. One has a {\it local} system
of functions $ \Omega _A$ on one-dimensional complex curve and the
corresponding set of parameters $
t_A$ so that it is possible to introduce a $ 1$-form in the space
with co-ordinates $ \left\{ \lambda,t \right\}$ where $\lambda $ is a
{\it local} parameter on a curve
\footnote{Actually it might be better to write
$$
\tilde{\omega} = \omega + \Omega _{\lambda } d\lambda
$$
and consider $ \Omega _{\lambda} = 1$ as a sort of ''gauge condition".}
\begin{equation}\label{1form}
\omega = \sum{\Omega _A \delta t_A}
\end{equation}
The Whitham equations are
\begin{equation}\label{gwhi}
\delta\omega\wedge\delta\omega = 0  \ \ \ \
\delta\omega = \partial _{\lambda} \Omega _A \delta\lambda\wedge\delta t_A +
\partial _B \Omega _A\delta t_A \wedge\delta t_B
\end{equation}
so that one needs to check the independent vanishing of the two
different terms -- $ \delta t ^4$ and $ \delta t^3\delta\lambda$.
The second term gives
\begin{equation}\label{3whi}
\sum{  \partial _\lambda \Omega _A\partial _B \Omega _C} = 0
\end{equation}
for the antisymmetrized sum or introducing some explicit co-ordinates
$ t_{A_0}\equiv x$ and
$ \Omega _{A_0}(z,t)\equiv p(z,t)$ is a local co-ordinate adjusted to
the fixed choice of parameters $ \left\{ t_A \right\}$
\be\label{pwhi}
\partial _A \Omega _B - \partial _B \Omega _A + \left\{ \Omega _A ,
\Omega _B \right\} = 0
\ \ \
\left\{ \Omega _A , \Omega _B \right\} \equiv {\partial \Omega _A \over
\partial x}{\partial \Omega _B \over \partial p} - {\partial \Omega
_B \over \partial x}{\partial \Omega _A \over \partial p}
\ee
In fact (\ref{pwhi}) strongly depend on the choice of the
local co-ordinate $ p$. The equations (\ref{gwhi}), (\ref{3whi}) and
(\ref{pwhi}) are defined only locally and have huge amount of solutions.

A possible way to restrict ourselves is to get ''globalized picture" related
with the ''modulation"
of parameters of the finite-gap solutions of integrable systems of KP-Toda type.
The KP $\tau$-function
associated with a given spectral Riemann curve is
\be\label{KPsol}
{\cal T}\{t_i\} = e^{\sum t_i\gamma _{ij}t_j}\vartheta\left({\bf \Phi}_0
+
\sum t_i{\bf k}_i\right) \ \ \
{\bf k}_i = \oint_{\bf B} d\Omega_i
\ee
where $ \vartheta$ is a Riemann theta-function and $d\Omega_i$ are
meromorphic 1-differentials with poles
of the order $i+1$ at a marked point $z_0$. They are completely
specified by the normalization relations
\be
\oint_{\bf A} d\Omega_i = 0
\label{norA}
\ee
and the asymptotic behaiviour
\be
d\Omega_i = \left(\xi^{-i-1} + o(\xi)\right) d\xi
\label{norc}
\ee
where $\xi$ is a local coordinate in the vicinity of $z_0$.
The moduli $\{ u_\alpha \}$ of spectral curve are
invariants of KP flows,
\be
\frac{\partial u_\alpha}{\partial t_i} = 0,
\ee
The way the moduli
depend on $t_i$ after the ''modulation" is defined  by the
Whitham equations (\ref{pwhi}). For $ z = z(\lambda ,t)$ so that
$\partial _i z = \{\Omega _i, z\}$ they acquire more simple form
\be\label{zwhi}
\frac{\partial d\Omega_i(z)}{\partial t_j} =
\frac{\partial d\Omega_j(z)}{\partial t_i}.
\ee
and imply that
\be\label{SS}
d\Omega_i(z) = \frac{\partial dS(z)}{\partial t_i}
\ee
and the equations for moduli, following from (\ref{zwhi}), are:
\be\label{whiv}
\frac{\partial u_\alpha}{\partial t_i} =
v_{ij}^{\alpha\beta}(u)\frac{\partial u_\beta}{\partial t_j}
\ee
with some (in general complicated) functions $v_{ij}^{\alpha\beta}$.

In the KdV (and Toda-chain) case  all the spectral curves are
hyperelliptic, and for the KdV $i$ takes only odd
values $i = 2j+1$, and
\be
d\Omega_{2j+1}(z) = \frac{{\cal P}_{j+g}(z)}{y(z)}dz,
\ee
the coefficients of the polynomials ${\cal P}_j$ being fixed
by normalization conditions (\ref{norA}), (\ref{norc})
(one usualy takes $z_0 = \infty$ and the local parameter
in the vicinity of this point is $\xi = z^{-1/2}$). In this case the
equations (\ref{whiv}) can be diagonalized if the co-ordinates
$ \left\{  u_\alpha \right\}$ on the moduli space are taken to be the
ramification points:
\begin{equation}\label{whihy}
v_{ij}^{\alpha\beta}(u) = \delta ^{\alpha\beta}\left.{d\Omega
_i(z)\over d\Omega _j(z)}\right| _{z=u_\alpha}
\end{equation}
Now, an important point is that the differential $dS(z)$ (\ref{SS}) can be
constructed for a generic finite-gap solution, i.e. for a solution determined
by a (spectral) complex curve \cite{DKN}. In general, it is a {\it meromorphic}
differential whose singularities are determined by physical properties
of the theory. Below several known examples
of exact solutions will be discussed and the corresponding generating
differentials (\ref{SS}) will be presented.

\section{Spherical solutions and topological gravity}

Let us, first, turn to the simplest nontrivial solutions of the systems
(\ref{pwhi}), (\ref{zwhi}) related with $2D$ topological string theories
given at least locally by the formulas
\begin{equation}\label{sphere}
z(\lambda ,t)^p = \lambda ^p + u_{p-2}(t)\lambda ^{p-2} + ... + u_0(t)
\end{equation}
so that
\be\label{spham}
\Omega _i(\lambda ,t) = z(\lambda ,t)^i_+
\ \ \
\Omega _i(z,t) = z^i + O(z^{{i \over n}-1})
\ee
and in the $ p=2$ example
\be\label{p2}
z^2 = \lambda ^2 + U(x,t)
\ \ \
\Omega _i(\lambda ,t) = z(\lambda ,t)^i_{+} =
\left( \lambda ^2 + U(x,t) \right)^{1 \over 2}_+
\nn \\
\Omega _i(z,t) = z^i + O(z^{{i \over 2}-1})
\ee
Now, in the simplest ''global picture" it is possible to interpret
$\lambda $ and $ z$ as global co-ordinates on sphere
with one marked point (where $\lambda = z = \infty$).

For particular choice of parameters $ \left\{ t_k \right\}$
this generic picture gives
Whitham solutions coming from the KP/Toda equations in the following way.
Starting with the ''zero-gap" solution to the KdV hierarchy
\be\label{0gap}
U(x,t) = u = const
\ \ \ \
\Psi (\lambda ,t) = e^{\sum_{k>0}{t_k z^k(\lambda )_+}}
\nn \\
\left( \partial _{t_1}^2 + u \right)\Psi = z^2\Psi
\ \ \ \
\partial _{t_k}\Psi  = \left( \partial _{t_1}^2 + u \right)^{k\over 2}_{+}\Psi
\ee
one comes by
\be\label{average}
\Omega _1 = \overline{\log\Psi (\lambda ,t)_{t_1}} =
z(\lambda )_{+} = \lambda = \sqrt{z^2 - u} \equiv p(z)
\nn \\
\Omega _i = \overline{\log\Psi (\lambda ,t)_{t_i}} = z^{i}(\lambda )_{+} =
\lambda ^3 + {3\over 2}u\lambda = \left( z^2 - u\right)^{3\over 2} +
{3\over 2}u\sqrt{z^2 - u}
\ee
to the formulas (\ref{p2}). The generating differential
(\ref{SS}) is now
\be\label{SKon}
dS(z) = \sum t_k d\Omega _k (z)
\ee
and the solution can be found in terms of the ''periods"
\be\label{resinf}
t_k = {1\over k}\res _{\infty}\left( z^{-k}dS\right)
\ee
For example the third KdV flow gives rise to
\beq\label{whisph}
\partial _{t_3}\Omega _1 (z) = \partial _{t_1}\Omega _3(z)
\eeq
which is equivalent to the Hopf or dispersionless KdV equation
\beq\label{hopf}
u_{t_3} -{3\over 2} uu_{t_1} = 0
\eeq
with generic solution
\be\label{hsph}
t_1 + {3\over 2}t_3u + P(u) = 0
\ \ \ \ \
P(u) \sim \sum t_{2n+1}u^n
\ee
Now, for the purposes of quantum field theory one needs to take from
(\ref{hsph}) the solution of the Whitham equation which is exact solution of
the ''full" KdV
\beq\label{x/t}
u = -{2\over 3}{t_1\over t_3}
\eeq
The generating differential (\ref{SS}) is now
\be\label{dstopgr}
\left. dS \right|_{t_3 = {2\over 3}, t_{k\neq 3} = 0} =
\lambda d\left( z^2 (\lambda )\right) = - (\lambda ^2 + u)d\lambda + d(\dots )
\ee
Then the solution to the linear problem
\beq\label{basph} \left(\partial _{t_1}^2 - {2\over 3}{t_1\over t_3}\right)\Psi
= z^2\Psi
\eeq
gives
\be\label{ai}
\left.\Psi (t_1,z)\right|_{t_3 = {2\over 3}} = Ai \left(
t_1 + z^2 \right)
\nn \\
\phi _i (z) \sim {\partial ^{i-1}\Psi
(z,t)\over\partial t_1^{i-1}} = \sqrt{2z}e^{-{2\over 3}z^3}\int dx
x^{i-1}e^{-{x^3\over 3} + xz^2}
\ee
and determinant formula \cite{mamo}
\beq\label{det}
{{\cal T} (t+T)\over{\cal T} (t)} = {\det \phi _i (z_j)\over \Delta (z)}
\eeq
results
in the $\tau $-function of the whole hierarchy in Miwa co-ordinates $T_k = -
{1\over k}\sum z_j^{-k}$.  The decomposition of $\log{\cal T} (T)$ gives
the correlators (\ref{corr}) of two-dimensional topological gravity.

\section{Elliptic curves and the Gurevich-Pitaevsky solutions}

Now let us demonstrate that the similiar consideration of the global solutions
related with higher genus Riemann surfaces (even in the elliptic case) gives
rise to physically more interesting solutions.
In contrast to the previous example Whitham times will be nontrivially
related to the moduli of the curve.
For example, from a standard definition (see \cite{Kri}) in case of the
elliptic curve $y^2 = \prod _{i=1}^3(x - u_i)$ one has
\beq\label{times}
t_k = {2\over k(2-k)}\res (x^{1-{k\over 2}}dy)
\eeq
or substituting $x = \wp (\xi ) + c$ and $y = {1\over
2}\wp '(\xi )$, where $\wp (\xi )$ is the Weierstrass $\wp$-function,
\be\label{times'}
t_k = -{2\over k(2-k)}\res \left( (\wp (\xi ) + c)^{1-{k\over 2}}
\wp ''(\xi )d\xi\right)
= \nn \\
= -{6\over k(2-k)}Res {d\xi\over \xi ^{6-k}}\left( 1+c\xi ^2 + {g_2\over
20}\xi ^4 +
...\right)^{1-{k\over 2}}\left( 1+ {g_2\over 60}\xi ^4 + ...\right) =
\nn \\
= {2\over 5}\delta _{k,5} - c\delta _{k,3} +\left( {3\over 4}c^2 -
{1\over
4}g_2\right)\delta _{k,1} + {\cal O}(t_{-k})
\ee
with dependent negative times $t_{-k} = t_{-k}(t_1,t_3)$ \cite{LGGKM}.

The elliptic (one-gap) solution to the KdV is
\be\label{ukdvsol}
U(t_1,t_3,\ldots| u) =
\frac{\partial^2}{\partial t_1^2} \log{\cal T} (t_1,t_3,\ldots|u) =
\\
= U_0 \wp (k_1t_1 + k_3t_3 + \ldots + \Phi_0 |\omega , \omega ') +
 {u\over 3}
\ee
and
\be
dp \equiv d\Omega_1(z) = \frac{z - \alpha (u)}{y(z)}dz, \nn \\
dQ \equiv d\Omega_3(z) = \frac{z^2 - \frac{1}{2}uz - \beta
(u)}{y(z)}dz.
\label{pp}
\ee
Normalization conditions (\ref{norA}) prescribe that
\be
\alpha (u) = \frac{\oint_A \frac{zdz}{y(z)}}
{\oint_A \frac{dz}{y(z)}}\ \ \ \ \
\beta (u) = \frac{\oint_A \frac{(z^2 - \frac{1}{2}uz)dz}{y(z)}}
{\oint_A \frac{dz}{y(z)}}.
\ee
The simplest elliptic example is the first Gurevich-Pitaevsky (GP) solution
\cite{GP} with the underlying spectral curve
\beq\label{elc1}
y^2 = (z^2 - 1)(z-u)
\eeq
specified by a requirement that all branching points except for $z = u$ are
{\it fixed} and do not obey Whitham deformation. The generating differential
corresponding to (\ref{elc1}) is given by \cite{WSGP}
\be\label{S}
dS(z) = \left(t_1 + t_3(z+\frac{1}{2}u)  + \ldots \right)
\frac{z-u}{y(z)}dz \ \stackreb{\{ t_{k>1}=0 \} }{=} \
t_1 \frac{z-u}{y(z)}dz
\ee
and it produces the simplest solution to (\ref{zwhi}) coming from the elliptic
curve. From (\ref{S}) one derives:
\be
\frac{\partial dS(z)}{\partial t_1} =
\left( z - u - (\frac{1}{2}t_1 + u t_3)
\frac{\partial u}{\partial t_1}\right)\frac{dz}{y(z)}, \nn \\
\frac{\partial dS(z)}{\partial t_3} =
\left( z^2 - \frac{1}{2}uz - \frac{1}{2}u^2 -
(\frac{1}{2}t_1 + u t_3)
\frac{\partial u}{\partial t_3}\right)\frac{dz}{y(z)}, \nn \\
\ldots ,
\ee
and comparison with explicit expressions (\ref{pp}) implies:
\be
(\frac{1}{2}t_1 + u t_3)\frac{\partial u}{\partial t_1} =
\alpha (u) - u, \nn \\
(\frac{1}{2}t_1 + u t_3)\frac{\partial u}{\partial t_3} =
\beta (u) - \frac{1}{2}u^2.
\label{preW}
\ee
In other words, this construction provides the first GP solution to the
Whitham equation
\be\label{GPe}
\frac{\partial u}{\partial t_3} = v_{31}(u)
\frac{\partial u}{\partial t_1},
\ee
with
\be
v_{31}(u) = \frac{\beta(u)-\frac{1}{2}u^2}{\alpha(u)-u} =
\left.\frac{d\Omega_3(z)}{d\Omega_1(z)}\right|_{z=u},
\ee
which can be expressed through elliptic integrals \cite{GP}.

The elliptic solution with all moving branch points is given instead
of (\ref{S}) by
\be\label{2gps}
y^2 = \prod _{i=1}^3 (z-u_i)
\ \ \ \ \
dS = y(z)dz
\ee
and all $u_i$ are some functions of the Whitham times. Then one has to
fulfil
\be
dp = {\partial\over\partial t_1}dS = ydz\sum _{i=1}^3{v_i\over z-u_i}=
{z-\alpha (u)\over y}dz
\nn \\
dQ = {\partial\over\partial t_3}dS = ydz\sum _{i=1}^3{w_i\over z-u_i}=
{z^2 - \left( {1\over 2}\sum u_i\right)z -\beta (u)\over y}dz
\ee
where
\be
v_i = -{1\over 2}{\partial u_i\over\partial t_1} \nn \\
w_i = -{1\over 2}{\partial u_i\over\partial t_3}
\ee
and $\alpha (u)$ and $\beta (u)$ are defined as before. Now what one gets
is a simple linear system of the equations
\be
R_{ij}(u)v_j = V_i \ \ \ \
R_{ij}(u)w_j = W_j
\nn \\
R_{ij}(u) = \left(
\begin{array}{ccc}
    1     &     1     &     1     \\
u_2 + u_3 & u_1 + u_3 & u_1 + u_2 \\
  u_2u_3  &   u_1u_3  &   u_1u_2
\end{array}\right)
\nn \\
V^{T} = \left( 0, -1, -\alpha (u)\right)
\ \ \ \
W^{T} = (1, {1\over 2}\left( u_1 + u_2 + u_3), -\beta (u)\right)
\ee
The solution of linear system looks similiar to (\ref{GPe}), for example:
\beq
{w_1\over v_1} = {u_1^2 - {1\over 2}(u_1 + u_2 + u_3) - \beta (u)
\over u_1 - \alpha (u)}
\eeq
and this class of solutions contain, for example, pure (non-topological!)
$2D$ gravity. The second GP solution
\footnote{corresponding to the Yang-Lee edge singularity or $(2,5)$ conformal
minimal model interacting with $2D$ gravity}
\be\label{GP2}
y^2 = \prod _{i=1}^3 (z - u_i)
\ \ \ \ \
dS = (z - e)ydz
\ee
can be obtained from elliptic curve with a marked point.
In all these cases, one might expect that a ''modulated" $\tau $-function
(cf. with (\ref{KPsol})) would still have a form
\beq\label{taudef}
{\cal T}\{t_i\} = e^{{\cal F}(t)}\vartheta
\left( {\bf S} (t) \right)
\eeq
so that
\beq
{\bf k}_i(t) = {\partial {\bf S}\over\partial t_i}
\eeq
and the poles of the ''effective" potential
\beq
u(t) = {\partial ^2 \over \partial t_1^2}\log {\cal T}
\eeq
can be identified with the massive excitations (\ref{mass}).

\section{Toda-type curves}

The above procedure can be also applied to the finite-gap solutions
determined by more general types of algebraic curves. The special case
of the Toda-chain curves \cite{Kri/Dub}
\be\label{todacurv}
y^2 = P_n(z)^2 - 1
\ \ \ \ \
2P_n(z) = w + w^{-1} \ \ \ \ \ 2y = w - w^{-1}
\ee
was notices \cite{WSGP,MartWa,Tak} to correspond to the $N=2$ SYM exact
solutions with the gauge group $SU(n = N_c)$
\footnote{ In fact one can show that the elliptic curve (\ref{elc1}) is a
particular case in the class of curves discussed below. Indeed if one takes
$$
w + w^{-1} = 2P_1(z) = 2z
$$
and considers the double cover
$$
z = u + x^2
$$
one gets (\ref{elc1}) after the change of variables $y_{new} = yx$
\cite{WiDo,IM}.}.
The generating functional for the Toda-type theories is given by
\be\label{todaS}
dS = dz\log \left(P_n(z) + \sqrt{P_n(z)^2 - 1}\right)
\nn \\
d\Omega _j (z) = {\partial dS\over\partial t_j} = {dz\over\sqrt{P_n(z)^2 - 1}}
{\partial P_n(z)\over\partial t_j}
\ee
and again corresponds to a specific Whitham solution. Indeed, for the $SL(2)$
case the generating differential (\ref{S}) had a simple zero only in one of
the branch points $z = u$, which means that the only branch point is influenced
by the Whitham evolution. Now, it is easy to see that the zeroes of generating
differential are at the points
\beq\label{zertS}
P_n(z) = 1
\eeq
i.e. the Whitham dynamics is nontrivial only for the half of the branch points
($y = 0$ in (\ref{todacurv})). In the special case of $n=1$ after the change
of variables discussed in the footnote above the generating differential
(\ref{todaS}) acquires the form of (\ref{S}).

Now let us turn to the question what are the most general possible type
of an integrable system appearing in the formulation of the exact quantum
field theory solution. The pure Yang-Mills theory corresponds to the
Toda-chain (KdV) systems being the simplest ones among integrable models
of this type. Their existing generalizations can be possible looked for
in two different directions. One and the most obvious is towards the
Calogero-Moser particle systems and this is the case discussed from
various points of view in \cite{WiDo,GM,IM,Mart}. This line ends up at
the free $N=4$ supersymmetric Yang-Mills theory which is described in
terms of {\it holomorphic} generating differential $dS$
\footnote{The fact that the $N=4$ SYM theories are described in terms of
holomorphic differentials can be easily understood assuming the connection of
the Whitham hierarchy with the renormalization group approach. Indeed, on
one hand the $N=4$ theory has zero $\beta $-functions and on the other
hand it correspond to a trivial Whitham modulation. Since the holomorphic
differentials have no singularities on spectral curve there is no natural
way to introduce higher times of the KP/Toda type which is again consistent
with the conformal invariance or trivial interaction with the ''scale part"
of metrics in $N=4$ theory.
}.

More interesting question is coupling of the $N=2$ SUSY YM theory with
{\it real} matter, i.e.
belonging to the fundamental representation of the gauge group. Such theories
are supposed to be described by the following curves \cite{WS,HO}
\be\label{matter}
y^2 = P_n(z)^2 - P_m(z)
\ee
where $m = N_f$ is the number of flavours or matter multiplets. These curves
correspond to rather specific integrable systems such as Goryachev-Chaplygin
top for $n = 3$ and massless $m=2$. An integrable system corresponding
to a generic curve can be
constructed by standard methods of finite-gap integration \cite{DKN}.

A generic integrable system of this type is determined by a (genus $g$)
spectral curve, a {\it meromorphic} generating differential $dS$ and a set of
$g$ given points $\{ P_1,\dots,P_g\} $ or initial conditions. The symplectic
structure is defined by a (closed) form
\be\label{symp}
\Omega = \sum _{\alpha = 1}^g  {\partial\over\partial u_{\alpha}}\left.\left(
{dS\over dz}  \right)\right|_{z_{\alpha} = z(P_{\alpha})} \delta u_{\alpha}
\wedge \delta z_{\alpha}
\ee
where as usual the moduli $\{ u_{\alpha}\} $ for the hyperelliptic cases
(\ref{2gps}), (\ref{GP2}), (\ref{todacurv}) and (\ref{matter}) can be taken
as co-ordinates of the branch points.
Any two functions of moduli $\{ u_{\alpha} \} $ commute
with respect to the Poisson bracket induced by (\ref{symp}) thus giving
the set of $g$ independent Hamiltonians. The integrals
\be
\theta _{\alpha } = \sum_{\beta = 1}^g \int _{z_0}^{z_{\alpha}}
{\partial dS\over \partial u_{\beta}}
\ee
are the co-ordinates on the Jacobian of a spectral curve if
${\partial dS\over \partial u_{\beta}} = d\omega _{\beta }$ are the
holomorphic differentials.

The spectral curves (\ref{matter}) can be identically described by the equation
\be\label{monmat}
\det \left( T(z) - w\right) = w^2 - \Tr T(z) w + \det T(z) = 0
\ \ \ \
2y = w - w^{-1}
\ee
where $T(z)$ is a monodromy matrix. The integrable system can be restored by
constructing the Baker-Akhiezer function
on a curve with marked points at the singularities of the monodromy matrix
and the integrability can be proved using for example $r$-matrix formalism.
In cases of (\ref{todacurv}), (\ref{matter}) the singular points
are $P_{\pm} = (\infty , \pm )$. The modulation of the integrals of motion
again gives rise to the Whitham equations generated by
\be\label{Smat}
dS = dz\left( \log \left(P_n(z) + \sqrt{P_n(z)^2 - P_m(z)}\right) -
{1\over 2}\log P_m(z) \right)
\ee

\section{String equation. KdV case}

Finally, I will make few remarks concerning the way how in all the cases above
a particular solution to the Whitham equations was distinguished. The idea is
that ''physical" solution to the Whitham hierarchy obeys also some ''string
equation"
which has a clear interpretation in terms of the spectral function for the
auxiliary linear problem.

Let us consider the Lax equation for KdV
\beq\label{schrod}
\left(\partial _x^2 + U(x)\right)\Psi (E,x) = E\Psi (E,x)
\eeq
as a Schr{\" o}dinger equation. The partition function
\be\label{part}
Z(t) = \Tr e^{-t{\hat H}} = \Tr e^{-t(\partial _x^2 + U)}
= \int _{x(0)=x(t)}Dx e^{\int _0^t {\dot x}^2 + U(x)}
\ee
under the Laplace transform gives
\beq\label{rho}
\rho (E) = \int _0^{\infty}dt e^{-tE} Z(t) =
\Tr {1\over \partial _x^2 + U(x) - E } = \int dx\sum _{n\geq
0}{R_n[U(x)]\over E^{n+{1\over 2}}}
\eeq
Now, using
\beq\label{gd}
R_n[U(x)] = {\partial ^2\over\partial x\partial t_{2n-1}}\log{\cal T}
(x=t_1,t_3,...)
\eeq
one can rewrite (\ref{gd}) as
\be\label{rhotau}
\rho (E) = \sum _{n\geq 0}{1\over E^{n+{1\over 2}}}\int
dx
{\partial ^2\over\partial x\partial t_{2n-1}}\log{\cal T} \sim
\nn \\
\sim \sum _{n\geq 0}{1\over E^{n+{1\over 2}}}
{\partial \over\partial t_{2n-1}}\log{\cal T} (t_k - {1\over kE^{k\over 2}})
\ee
or
\be\label{rhoq}
\rho (E) = {\partial\over\partial E}\log{\cal T} (t_k -
{1\over kE^{k\over 2}}) \sim z^q
\ \ \  \
z = \sqrt{E}
\ee
where the behaiviour $z^q$ is provided by string equation (specific initial
condition),
i.e. the solutions to string equations are characterized by growing at
infinity spectral function and the order of the singularity at infinity
gives a particular critical point.

The sense
of the Whitham deformation is variation of the spectrum of
the auxiliary linear problem (\ref{schrod}).
In terms of the path integral for the ''spectral" fermions is
\be\label{ferpi}
\int D\psi ^{\ast }D\psi e^{\int \psi ^{\ast}\left(\partial _x^2
+ u(x) - E\right)\psi }
= \det\left( \partial _x^2 + u(x) - E\right)
\ee
using the relation
\beq\label{relat}
\log\det\left( \partial _x^2 + U(x) - E\right)  =
\Tr\log\left( \partial _x^2 + U(x) - E\right)
\eeq
and taking into account (\ref{rho}) one finds
\beq\label{rhorel}
{\partial\over\partial E}\log\det\left( \partial _x^2 + U(x) - E\right)  =
\Tr{1\over \partial _x^2 + U(x) - E} = \rho (E)
\eeq
After integration the fermionic effective action
\beq\label{ferea}
\log\det\left( \partial _x^2 + U(x) - E\right)  =
\int ^E d\epsilon\rho (\epsilon ) = S(E)
\eeq
coincides with the ''Whitham effective action" $S(E) = \int ^E dS $ in the
sense of \cite{Kri,KM}. The
total energy can be divided into fermionic and potential pieces
\beq\label{pei}
E_{tot} = \sum _{fermions}E + E_{pot}\left[ U(x) \right]
\eeq
with measure of evaluating the sum in (\ref{pei}) determined by
a quasimomentum so that
\be\label{feer}
E_{ferm} = \int Edp
\ee
Now, the local extremum of (\ref{pei}) corresponds to the choice of a certain
critical point if one takes
\be\label{krus}
E_{pot}\left[ U(x) \right] = \sum _{k\leq q} t_k \int dx R_k \left[ U(x)\right]
\ee
which gives a relation of the (\ref{times}), (\ref{times'}) type. Thus we see
that the quantum dynamics of the fermions entering the auxiliary linear
problem gives rise to the deformation of spectral curve and nontrivial dynamics
on its moduli space.

\section{Conclusion}

In this note I have tried to demonstrate the role played by the integrable
equations in the nonperturbative approach to quantum string and supersymmetric
gauge field
theories. Using the general phenomenon that all known exact nonperturbative
solutions can be described in terms of spectral curves and corresponding
solutions to integrable systems one might hope that arising integrable
equations of hydrodynamical type can be interpreted in terms of generalization
of renormalization group beyond the perturbation theory.
Amazingly enough this construction does not depend too much upon the ''bare"
quantum field theory, thus the language of integrable systems seems to be
adequate and universal for studying nonperturbative phenomena.

\section{Acknowledgments}

I am deeply indebted to V.Fock, A.Gerasimov, A.Gurevich, S.Kharchev,
A.Mironov, N.Nekrasov, S.Novikov, A.Zabrodin and
especially to A.Gorsky, I.Krichever and A.Morozov for illuminating discussions.
I am grateful to J.Ambjorn and A.Niemi for warm hospitality at Niels Bohr
Institute and Institute of Theoretical Physics at Uppsala University where
this work has been started and to the organizers
of the conferences where these results were presented. The work was
partially supported by the
RFFI grant 95-01-01106, INTAS grant 93-0633 and by
NFR-grant No F-GF 06821-305 of the Swedish Natural Science Research Council.

\end{document}